\documentclass[12pt]{article}
\usepackage{amsfonts,graphicx,amsmath}

\newcommand{\be}{\begin{equation}}
\newcommand{\ee}{\end{equation}}
\newcommand{\bea}{\begin{eqnarray}}
\newcommand{\eea}{\end{eqnarray}}
\newcommand{\beas}{\begin{eqnarray*}}
\newcommand{\eeas}{\end{eqnarray*}}

\font\cmss=cmss12

\newcommand{\phinotp} [1]{{\phi_{0#1}^{\it Poincare}}}

\newcommand{\phinotrL}[1]{{\phi_{0#1}^{\it Rindler,\,L}}}
\newcommand{\phinotrR}[1]{{\phi_{0#1}^{\it Rindler,\,R}}}

\newcommand{\hw}{\hat{\omega}}
\newcommand{\hk}{\hat{k}}
\newcommand{\htt}{\hat{t}}
\newcommand{\hp}{\hat{\phi}}

\begin{document}

\baselineskip 14 pt
\parskip 12 pt

\begin{titlepage}
\begin{flushright}
{\small CU-TP-1162} \\
{\small hep-th/0612053}
\end{flushright}

\vspace{1mm}

\begin{center}

{\Large \bf Local bulk operators in AdS/CFT: A} \\
{\Large \bf holographic description of the black hole interior}

\vspace{4mm}

Alex Hamilton,${}^1$ Daniel Kabat,${}^1$ Gilad Lifschytz,${}^2$
and David A.\ Lowe${}^3$

\vspace{2mm}

${}^1${\small \sl Department of Physics} \\
{\small \sl Columbia University, New York, NY 10027 USA} \\
{\small \tt hamilton, kabat@phys.columbia.edu}

\vspace{1mm}

${}^2${\small \sl Department of Mathematics and Physics and CCMSC} \\
{\small \sl University of Haifa at Oranim, Tivon 36006 ISRAEL} \\
{\small \tt giladl@research.haifa.ac.il}

\vspace{1mm}

${}^3${\small \sl Department of Physics} \\
{\small \sl Brown University, Providence, RI 02912 USA} \\
{\small \tt lowe@brown.edu}

\end{center}

\vspace{2mm}

\noindent
To gain insight into how bulk locality emerges from the holographic
conformal field theory, we reformulate the bulk to boundary map in as
local a way as possible.  In previous work, we carried out this
program for Lorentzian AdS, and showed the support on the boundary
could always be reduced to a compact region spacelike separated from
the bulk point. In the present work the idea is extended to a
complexified boundary, where spatial coordinates are continued to
imaginary values. This continuation enables us to represent a local
bulk operator as a CFT operator with support on a finite disc on the
complexified boundary.  We treat general AdS in Poincar\'e coordinates
and AdS${}_3$ in Rindler coordinates.  We represent bulk operators
inside the horizon of a BTZ black hole and we verify that the correct
bulk two point functions are reproduced, including the divergence when
one point hits the BTZ singularity.  We comment on the holographic
description of black holes formed by collapse and discuss locality and
holographic entropy counting at finite $N$.

\end{titlepage}

\section{Introduction}\label{Intro}

The AdS/CFT correspondence relates string theory in an asymptotically
anti-de Sitter (AdS) background to a conformal field theory (CFT) living on the
boundary of AdS \cite{Maldacena:1997re,Gubser:1998bc,Witten:1998qj,Aharony:1999ti}.
The main observable of interest in the original work on AdS/CFT was the boundary S-matrix. In the present work we will focus instead on how one might recover approximate local bulk quantities from the CFT. Even with interactions included, one can hope to recover quasi-local observables in the gravitational theory \cite{Giddings:2005id}. In this paper we study this in detail, generalizing our earlier work \cite{hkll, hkll2}. We develop the map at leading order in $1/N$ where we can treat the gravity theory semiclassically and work with free scalar fields.   

The original bulk to boundary map of Euclidean AdS \cite{Gubser:1998bc,Witten:1998qj} and its Lorentzian generalization \cite{Balasubramanian:1998de,Balasubramanian:1998sn} has been reformulated and studied further in \cite{Banks:1998dd,Balasubramanian:1999ri,bena,Susskind:1998vk,Polchinski:1999ry}.  In these works, one can view the construction of a local bulk operator as an integral  over the entire boundary of AdS.  Vanishing of commutators of local bulk operators at spacelike separation relies on delicate cancellations in this approach \cite{bena}. The strategy we will adopt, following our earlier work \cite{hkll,hkll2}, is to reformulate the bulk to boundary map so that the support
on the boundary is as small as possible.

It is worth emphasizing the physical relevance of our approach.  By representing bulk operators as operators on the boundary
with compact support -- in fact with support that is
as small as possible -- we can have bulk operators whose
dual boundary operators are spacelike separated.  Such bulk operators
will manifestly commute with each other just by locality of the
boundary theory.  This statement will continue to hold at finite $N$.  Moreover we will find interesting applications of this basis of operators to the study of black hole interiors and singularities, as well as holographic entropy counting.

We will use the following framework developed in \cite{hkll,hkll2}. The first of these works \cite{hkll} mainly considered two-dimensional AdS space, and showed the boundary support of a local bulk operator could be reduced to points spacelike separated from the bulk point. This was generalized to the higher dimensional case in \cite{hkll2}.
In Lorentzian AdS, a free bulk scalar field $\phi$
is dual to a
non-local operator in the CFT, via a correspondence
\be
\label{BulkDuality}
\phi(x,Z) \leftrightarrow \int dx' \, K(x' \vert x,Z) {\cal O}(x')\,.
\ee
Here $Z$ is a radial coordinate in AdS which vanishes at the boundary, $x$ represents coordinates
along the boundary, and ${\cal O}(x')$ is a local operator
in the CFT. A similar approach has been considered previously in \cite{Banks:1998dd,Balasubramanian:1999ri,bena,Polchinski:1999ry}.  We refer to the kernel $K$ as a smearing function.  This correspondence
can be used inside correlation functions, for example
\[
\langle \phi(x_1,Z_1) \phi(x_2,Z_2) \rangle_{SUGRA} = \int dx_1' dx_2' \, K(x_1' \vert x_1,Z_1)
K(x_2' \vert x_2,Z_2) \langle {\cal O}(x_1') {\cal O}(x_2') \rangle_{CFT}\,.
\]

To construct smearing functions one begins with a field in Lorentzian AdS that satisfies the
free wave equation and has normalizable fall-off near the boundary of AdS,\footnote{This is to be contrasted with the original formulations of the bulk to boundary map \cite{Balasubramanian:1998de,Balasubramanian:1998sn} that include both normalizable and non-normalizable modes on the boundary.}
\[
\phi(x,Z) \sim Z^\Delta \phi_0(x) \qquad \hbox{\rm as $Z \rightarrow 0$}\,.
\]
The parameter $\Delta$ is related to the mass of the field.
The boundary field $\phi_0(x)$ is dual to a local operator in the CFT\footnote{Again this should be compared with the original formulation of the bulk to boundary map where the non-normalizable component at the boundary is dual to a source for the CFT operator.}
\be
\label{LocalDuality}
\phi_0(x) \leftrightarrow {\cal O}(x)~.
\ee
We will construct smearing functions that let us solve for the bulk field in terms of the boundary field
\be
\phi(x,Z) = \int dx' \, K(x' \vert x,Z) \phi_0(x')~.
\label{holo}
\ee
$K$ should not be confused with the standard bulk-to-boundary propagator \cite{Witten:1998qj},
since our smearing functions generate {\em normalizable} solutions to the {\em Lorentzian}
equations of motion.  Using the duality (\ref{LocalDuality}),
we obtain the relation between bulk and boundary operators given in (\ref{BulkDuality}).

Solving for
the bulk field in terms of the boundary field is not a standard Cauchy problem: since the ``initial conditions''
are specified on a timelike hypersurface we have neither existence nor uniqueness theorems.  In global AdS it
was shown that, although the smearing function is not unique, one can always construct a smearing function which
has support on the boundary at points which are spacelike separated from the bulk point \cite{hkll,hkll2}. 
It is then interesting to see if a stronger statement can be made.  Can we further reduce the support on the boundary?
This was studied in \cite{hkll2}, where smearing functions for AdS${}_3$ were constructed in accelerating Rindler coordinates.
It was shown that smearing functions can only be constructed by analytically continuing the boundary coordinates to complex values,
since the naive expression derived from mode sums was divergent.
This continuation leads to a well-defined smearing function with compact
support on the complexified boundary of the Rindler patch; it can be thought of as arising from a retarded
Green's function in de Sitter space.  Moreover the support shrinks to a point as the bulk
point approaches the boundary.  In this way we recover the expected relation (\ref{LocalDuality}).

It thus seems the most economical description of local bulk physics in AdS/CFT requires the use of
complexified boundary coordinates.  Complexified coordinates also appeared
in \cite{Balasubramanian:2003kq}, and have been used to study the region inside horizons in \cite{Kraus:2002iv,Fidkowski:2003nf,Festuccia:2005pi,Festuccia:2006sa,Freivogel:2005qh}.
For other approaches to recovering bulk physics see \cite{Hubeny:2006yu,Hammersley:2006cp,Yang:2006pc}.

An outline of this paper is as follows.
In section \ref{Poincare}  we extend the work of \cite{hkll2} and use complexified boundary coordinates to construct compact
smearing functions in AdS spacetimes of general dimension in two ways.  First we work in Poincar\'e coordinates and perform a
Poincar\'e mode sum, then we Wick rotate to de Sitter space and use
a retarded Green's function. In section \ref{AdS3} we
translate our AdS${}_3$ results into Rindler coordinates and show that
we recover bulk correlators inside the Rindler horizon.  After these preliminaries we
develop applications of this new formulation of the bulk/boundary map to black holes and to holographic entropy counting.
In section \ref{BTZ}
we argue that the Rindler smearing functions can also be used in a BTZ spacetime and
we show how the BTZ singularity manifests itself in the conformal field theory.  In section \ref{collapse} we discuss
local operators inside the horizon of an AdS black hole formed by collapse, where there is only a single asymptotic
AdS region.  This provides evidence that our results will generalize to time-dependent situations. Finally in section \ref{locality} we explain how the 
number of degrees of freedom is reduced at finite $N$ and how this leads to a new perspective on holographic entropy counting.  

\section{Poincar\'e coordinates}\label{Poincare}

In this section we construct a compact smearing function for a
general-dimensional AdS spacetime.  We obtain the result in two
ways: by performing the Poincar\'e mode sum in section \ref{newpoincare},
and by Wick rotating to de Sitter space in section \ref{deSitter}.

\subsection{Preliminaries}

We will work in AdS${}_D$ in Poincar\'e coordinates, with metric
\be
\label{PoincareMetric}
ds^2 = {R^2 \over Z^2} \left(-dT^2 + \vert dX \vert^2 + dZ^2\right)\,.
\ee
Here $R$ is the AdS radius.  The coordinates range over $0 < Z < \infty$, $-\infty < T < \infty$, and
$X \in {\mathbb R}^{d-1}$ where $d = D-1$.  An AdS-invariant distance function is provided by
\be
\label{PoincareDistance}
\sigma(T,X,Z \vert T',X',Z') = {1 \over 2 Z Z'} \left(Z^2 + Z'{}^2 + \vert X - X' \vert^2 - (T - T')^2\right)\,.
\ee
We consider a free scalar field of mass $m$ in this background.  Normalizable
solutions to the free wave equation $(-\Box + m^2) \phi = 0$ can be expanded in a complete set of modes
\be
\label{PoincareModes}
\phi(T,X,Z) = \int_{\vert \omega \vert > \vert k \vert} \!\!\!\!\!\! d\omega d^{d-1}k \,
a_{\omega k} e^{-i \omega T} e^{i k \cdot X} Z^{d/2} J_\nu(\sqrt{\omega^2 - k^2} Z)~.
\ee
The Bessel function has order $\nu = \Delta - d / 2$ where
$\Delta = {d \over 2} + \sqrt{{d^2 \over 4} + m^2 R^2}$ is the conformal dimension of the
corresponding operator.

In Poincar\'e coordinates we define the boundary field by
\bea
\label{boundfield}
\phinotp{}(T,X) & = & \lim_{Z \rightarrow 0} {1 \over Z^\Delta} \phi(T,X,Z) \\
\nonumber
& = & {1 \over 2^\nu \Gamma(\nu + 1)} \int_{\vert \omega \vert > \vert k \vert} \!\!\!\!\!\! d\omega d^{d-1}k \,
a_{\omega k} e^{-i \omega T} e^{i k \cdot X} (\omega^2 - k^2)^{\nu/2}\,.
\eea
Note that
\be
a_{\omega k} = {2^\nu \Gamma(\nu + 1) \over (2 \pi)^d (\omega^2 - k^2)^{\nu/2}} \int dT d^{d-1}X \, e^{i \omega T}
e^{-i k \cdot X} \phinotp{}(T,X)\,.
\ee
Substituting this back into the bulk mode expansion (\ref{PoincareModes}), we obtain an
expression for the bulk field in terms of the boundary field, namely
\be
\label{PoincareSmear}
\phi(T,X,Z) = \int dT' d^{d-1}X' \, K(T',X' \vert T,X,Z) \phinotp{}(T',X')
\ee
where
\bea
\nonumber
K(T',X' \vert T,X,Z) & = & {2^\nu \Gamma(\nu + 1) \over (2 \pi)^d} \int_{\vert \omega \vert > \vert k \vert}
\!\!\!\!\!\! d\omega d^{d-1}k \, e^{-i \omega (T - T')} e^{i k \cdot (X - X')} \\
& & Z^{d/2} J_\nu(\sqrt{\omega^2 - k^2} Z) / (\omega^2 - k^2)^{\nu/2}\,.
\eea
One could proceed to evaluate this integral representation for $K$ along the lines of \cite{bena,hkll,hkll2}.  However one
generically obtains a smearing function with support on the entire boundary of the Poincar\'e
patch.\footnote{In even-dimensional AdS one can restrict to spacelike separation in the Poincar\'e
patch \cite{hkll,hkll2}.}  In the following we will improve on this by constructing smearing functions that make manifest the
property that local bulk operators go over to local boundary operators as the bulk point approaches the boundary.  

Such smearing functions require complexifying the boundary spatial coordinates $X$.
We will establish this in two ways: in section \ref{newpoincare}, for fields in AdS${}_3$, by starting with the mode sum
(\ref{PoincareSmear}) and performing a suitable analytic continuation, and again in section \ref{deSitter}, for fields in general dimensional AdS,
by Wick rotating to de Sitter space and using a retarded de Sitter Green's function.

\subsection{Poincar\'e mode sum}\label{newpoincare}

Consider a field in AdS${}_3$.  The Poincar\'e mode sum  (\ref{PoincareSmear}) reads
\beas
\phi(T,X,Z) & = & {2^\nu \Gamma(\nu + 1) \over 4 \pi^2} \int_{\vert \omega \vert > \vert k \vert} d\omega dk \,
{Z J_\nu(\sqrt{\omega^2 - k^2}Z) \over (\omega^2 - k^2)^{\nu/2}} \\
& & \times \left( \int dT' dX' \, e^{-i \omega (T-T')} e^{i k (X-X')} \phinotp{}(T',X') \right)
\eeas
The Poincar\'e boundary field has no Fourier components with $\vert
\omega \vert < \vert k \vert$, so provided we perform the $T'$ and $X'$
integrals first we can subsequently integrate over $\omega$ and $k$ without restriction.
Thus
\be
\label{fieldef}
\phi(T,X,Z) = 2^{\nu}\Gamma (\nu+1)\int d\omega dk \, e^{-i\omega T} e^{ikX}\frac{ZJ_{\nu}(\sqrt{\omega^{2}-k^{2}}Z)}
{(\omega^{2}-k^{2})^{\nu/2}}\phinotp{}(\omega,k)
\ee
where $\phinotp{}(\omega,k)$ is the Fourier transform of the boundary field.
We now use the two integrals
\begin{eqnarray}
&&\int_{0}^{2\pi}d\theta \, e^{-ir\omega \sin \theta -kr\cos \theta}=2\pi J_{0} 
(r\sqrt{\omega^{2}-k^{2}})\\
&&\int_{0}^{1}rdr\, (1-r^{2})^{\nu-1}J_{0}(br)=2^{\nu-1}\Gamma(\nu)b^{-\nu}J_{\nu}(b)
\end{eqnarray}
to obtain
\begin{equation}
\frac{J_{\nu}(\sqrt{\omega^{2}-k^{2}}Z)}{(\omega^{2}-k^{2})^{\nu/2}}=
{1 \over \pi (2Z)^{\nu} \Gamma(\nu)} \int_{T'{}^{2}+Y'{}^{2}<Z^2} \!\!\!\! dT' dY' \, (Z^{2}-T'{}^{2}-Y'{}^{2})^{\nu-1} e^{-i\omega T'}e^{-kY'}\,.
\end{equation}
Inserting this into (\ref{fieldef}) one gets
\begin{eqnarray}
\nonumber
\phi(T,X,Z) & = & \frac{\nu}{\pi}\int_{T'{}^{2}+Y'{}^{2}<Z^2}\!\!\!\!dT'dY'\,\left({Z^{2}-T'{}^{2}-Y'{}^{2}\over Z}\right)^{\nu-1}\\
\label{field2}
& & \qquad \int d\omega dk \, e^{-i\omega (T+T')}e^{ik(X+iY')}\phinotp{}(\omega,k)
\end{eqnarray}
We identify the second line of (\ref{field2}) as $\phinotp{}(T+T',X+iY')$, so we can write (recall $\nu = \Delta - 1$)
\begin{equation}
\label{PoincareModeResult}
\phi(T,X,Z)=\frac{\Delta - 1}{\pi}\int_{T'{}^2 + Y'{}^2 < Z^2}\!\!\!\!dT'dY'\,
\left(\frac{Z^{2}-T'{}^{2}-Y'{}^{2}}{Z}\right)^{\Delta - 2}\phinotp{}(T+T',X+iY')\,.
\end{equation}
That is, we've succeeded in expressing the bulk field in terms of an integral over a disk of radius $Z$ in the (real $T$,
imaginary $X$) plane.  We can express the result in terms of the invariant distance (\ref{PoincareDistance}),
\bea
\nonumber
\phi(T,X,Z) & = & {\Delta - 1 \over \pi} \int_{T'{}^2 + Y'{}^2 < Z^2}\!\!\!\!dT'dY'\,
\lim_{Z' \rightarrow 0} \big(2Z'\sigma(T,X,Z \vert T+T',X + iY',Z')\big)^{\Delta - 2} \\
& & \quad \phinotp{}(T+T',X+iY')
\eea
We'll obtain the generalization of this result to higher-dimensional AdS in the next subsection.

\subsection{de Sitter continuation}\label{deSitter}

\begin{figure}
\centerline{\includegraphics{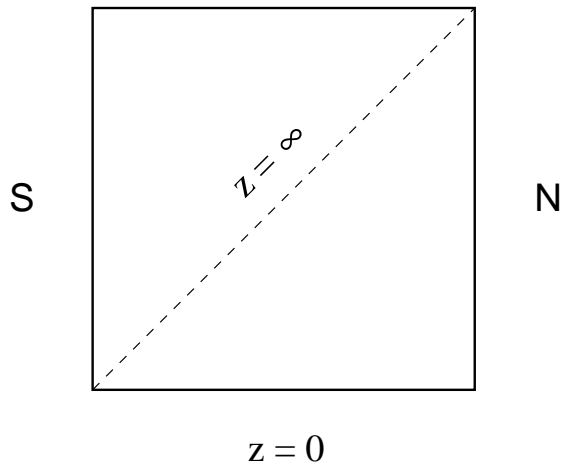}}
\caption{The Penrose diagram for de Sitter space.  Flat FRW coordinates cover the lower triangle.
Horizontal slices are spheres.  Each point on the diagram represents an $S^{d-1}$ which shrinks to a
point at the north and south poles (the right and left edges of the diagram).}
\label{P}
\end{figure}

Having seen that we need to analytically continue the boundary spatial
coordinates to complex values in order to obtain a smearing function with
compact support, we will now begin by Wick rotating the Poincar\'e longitudinal
spatial coordinates, setting $X = i Y$.  This turns the
AdS metric (\ref{PoincareMetric}) into
\[
ds^2 = {R^2 \over Z^2} \left(dZ^2 - dT^2 - \vert d Y \vert^2\right)\,.
\]
This is nothing but de Sitter space written in flat Friedmann-Robertson-Walker (FRW) coordinates,
with $Z$ playing the role of conformal time (note the flip in signature).
The AdS boundary becomes the past boundary of de Sitter
space.  Up to a divergent conformal factor the induced metric on the past
boundary is
\[
ds^2_{\rm bdy} = dT^2 + \vert d Y \vert^2
\]
i.e.\ a plane ${\mathbb R}^d$ which should be thought of as
compactified to a sphere $S^d$ by adding a point at infinity.  The Penrose diagram is shown in Fig.~\ref{P}.

In de Sitter space it's clear that the field at any point inside the
Poincar\'e patch can be expressed in terms of data on a compact region
of the past boundary.\footnote{To
go outside the Poincar\'e patch one would have to include the point at
infinity in ${\mathbb R}^d$.}  With this motivation we will construct a
retarded Green's function in de Sitter space and use it to reproduce and generalize the
smearing function (\ref{PoincareModeResult}) that we previously obtained from a Poincar\'e mode sum.

The de Sitter invariant distance
function is
\[
\sigma(T,Y,Z \vert T',Y',Z') = \frac{1}{2 Z Z'}\left(Z^2+Z'^2 - (T - T')^2 - \vert Y-Y'\vert^2 \right)~.
\]
We consider a scalar field of mass $m$ in de Sitter space.  For now we
take $m^2 R^2>1$, however later we will analytically continue $m^2 \to
- m^2$.  The analytically continued mass can be identified with the
mass of a field in AdS (note that the Wick rotation flips the
signature of the metric).

The field at some bulk point can be written in terms of the retarded
Green's function.  The retarded Green's function coincides with the imaginary part of
the commutator inside the past light-cone of the future point and
vanishes outside this region.  The field at some bulk point is
therefore
\be
\phi(T,Y,Z)=\int dT' d^{d-1}Y'\,\left(\frac{R}{Z'}\right)^{d-1}G_{\rm ret}(T,Y,Z \vert T',Y',Z')
\overleftrightarrow{\partial_{Z'}}\phi(T',Y',Z') \label{brfield}
\ee
where the region of integration is over a spacelike surface of fixed
$Z'$ inside the past light-cone of the bulk point.  In the $Z' \rightarrow 0$
limit this becomes the disk 
\be
\label{plc}
(T-T')^2 + \vert Y - Y' \vert^2 < Z^2\,.
\ee
As $Z'\to 0$ (with other coordinates held fixed) the retarded
Green's function takes the form \cite{Bousso:2001mw}
\[
G_{\rm ret}\sim i R^{-d+1}\left( c\left(-\sigma -i\epsilon\right)^{-d/2+i
\sqrt{m^{2}R^2-\left(\frac{d}{2}\right)^2}}+c^* \left(-\sigma-i\epsilon\right)^{-d/2-i
\sqrt{m^{2}R^2-\left(\frac{d}{2}\right)^2}}-c.c.\right)
\]
where we take branch cuts along the positive real $\sigma$ axis and where
\[
 c=  \frac{\Gamma\left(2 i\sqrt{m^2 R^2-\left(\frac{d}{2}\right)^2}\right) \Gamma\left(\frac{d}{2}-i\sqrt{m^2 R^2-\left(\frac{d}{2}\right)^2}\right) }{2^{-\frac{d}{2}+i
\sqrt{m^2 R^2-\left(\frac{d}{2}\right)^2} } \left(4\pi\right)^{\frac{d+1}{2}}  \Gamma\left(\frac{1}{2}+i\sqrt{m^2 R^2-\left(\frac{d}{2}\right)^2} \right) }~. 
\]
The boundary field is defined as usual using (\ref{boundfield}).
Choosing normalizable modes from the AdS viewpoint corresponds to
taking only positive frequencies in the $Z$ direction, which have a
$Z^{d/2+i\sqrt{m^2 R^2-\left(\frac{d}{2}\right)^2}}$ $Z$-dependence.

Evaluating (\ref{brfield}) as $Z' \to 0$ we obtain the
smearing function\footnote{Here we use the identities $\sin(\pi z) =
\frac{\pi}{\Gamma(z)\Gamma(1-z)}$ and
$\frac{\Gamma(2z)}{\Gamma(z)\Gamma(1/2+z)}=\frac{2^{2z-1}}{\sqrt{\pi}}$.}
\bea
K(T',Y' \vert T,Y,Z) & = & \frac{\Gamma \left( \Delta - \frac{d}{2} +1 \right) }{\pi^{d/2} \ \Gamma (\Delta - d + 1)}
\left(\frac{Z^2 - (T-T')^2 - \vert Y-Y'\vert^2}{Z} \right)^{\Delta - d} \nonumber \\*[5pt]
& & \quad \theta (Z^2-(T-T')^2-\vert Y-Y'\vert^2 )  \;. \label{smearedResult}
\eea

However at this point we still have $\Delta=\frac{d}{2}+i\sqrt{m^2 R^2-\left(\frac{d}{2}\right)^2}$.  By
analytically continuing $m^{2}\to-m^{2}$ we can take $\Delta$ to
coincide with the conformal dimension in AdS.  Since $\sigma>0$ in the
domain of integration this analytic continuation is straightforward. Furthermore we can shift $iY\to X+iY$ and $iY'\to X+iY'$,
assuming $\phinotp{}$ is analytic everywhere in the strip $|Y|<Z$; this is true for any given Poincar\'e mode function (\ref{PoincareModes}).
Thus we wind up with the integral representation
\bea
\nonumber
\phi(T,X,Z) & = & \frac{\Gamma \left( \Delta - \frac{d}{2} +1 \right) }{\pi^{d/2} \ \Gamma (\Delta - d + 1)}
\int_{T'{}^2 + \vert Y'\vert^2 < Z^2}\!\!\!\! dT' d^{d-1}Y' \, \left({Z^2 - T'{}^2 - \vert Y' \vert^2 \over Z}\right)^{\Delta - d} \\*[5pt]
& & \qquad \qquad \phinotp{}(T + T', X + iY')
\eea
This matches (\ref{PoincareModeResult}) for $d = 2$.
As a further check we can examine the limit $Z \to 0$ where we should recover (\ref{boundfield}). In this limit
the region of integration becomes very small so we can bring the boundary field out of the integral and we indeed
recover (\ref{boundfield}).

\subsection{Recovering bulk correlators}\label{correlators}

In this section we show that the smearing functions we have constructed can be used to reproduce
bulk correlation functions.  As a corollary, this shows that the operators we have defined will commute
with each other at bulk spacelike separation.  For simplicity we will only treat the case of a massless field in AdS${}_3$.

The Wightman function for a massless scalar in AdS${}_3$ is
\be
\label{AdS3Wightman}
G(x \vert x') = \langle 0 \vert \phi(x) \phi(x') \vert 0 \rangle_{SUGRA} = {1 \over 4 \pi R}
{1 \over \sqrt{\sigma^2 - 1} \, (\sigma + \sqrt{\sigma^2 - 1}\,)}
\ee
where $\sigma$ is defined in (\ref{PoincareDistance}), and where branch cuts are handled with a $T \rightarrow T - i \epsilon$
prescription.\footnote{This Wightman function identifies $\vert 0 \rangle$ as the Poincar\'e vacuum state.}  We'll consider
the correlation function between an arbitrary bulk point $(T,X,Z)$ and a point near the boundary with coordinates $(T'=0,X'=0,Z'
\rightarrow 0)$.  Taking the appropriate limit of (\ref{AdS3Wightman}) we have
\be
\label{ToGet}
\langle \phi(T,X,Z) \phinotp{}(0,0) \rangle_{SUGRA} = {1 \over 2 \pi R} \, {Z^2 \over (T^2 - X^2 - Z^2)^2}\,.
\ee
We'd like to reproduce this from the CFT.
To do this note that from (\ref{PoincareModeResult}) we have
\be
\phi(T,X,Z) = {1 \over \pi} \int_{T'{}^2 + Y'{}^2 < Z^2} \!\!\!\! dT' dY' \phinotp{}(T + T', X + i Y')\,.
\ee
Also by sending both points to the boundary in (\ref{AdS3Wightman}) we have the boundary correlator\footnote{We obtained this
from a boundary correlator in supergravity, but the result matches the correlator of local operators in the CFT.}
\be
\label{PoincareBdyCorr}
\langle \phinotp{}(T,X) \phinotp{}(0,0) \rangle_{CFT} = {1 \over 2 \pi R} \, {1 \over (T^2 - X^2)^2}\,.
\ee
Thus our claim is that we can reproduce (\ref{ToGet}) by computing
\bea
\nonumber
& & {1 \over \pi} \int_{T'{}^2 + Y'{}^2 < Z^2} \!\!\!\! dT' dY' \, \langle\phinotp{}(T + T', X + i Y')\phinotp{}(0,0)\rangle \\
\label{IntRep}
& = & {1 \over 2\pi^2R} \int_{T'{}^2 + Y'{}^2 < Z^2} \!\!\!\! dT' dY' \, {1 \over \big((T + T')^2 - (X + iY')^2\big)^2}
\eea
Let's begin by studying this in the regime
\be
\label{PoincareRange}
\vert T + X \vert > Z \quad {\rm and} \quad \vert T - X \vert > Z\,.
\ee
In this regime there are no poles in the range of integration, so (\ref{IntRep}) is well-defined without having to give a prescription
for dealing with light-cone singularities in the CFT.  It's convenient to work in polar coordinates, setting $T' = r \cos \theta$ and
$Y' = r \sin \theta$.  Defining $z=e^{i\theta}$ we have
\begin{equation}
\label{PoincareContour}
{1 \over \pi R} \int_0^Z rdr \oint_{|z|=1} {dz \over 2 \pi i} \, \frac{z}{(T + X + rz)^2 (z(T - X) + r)^2}
\end{equation}
Evaluating the contour integral gives
\begin{equation}
{1 \over \pi R} \int_0^Z rdr \frac{T^2-X^2+r^2}{(T^2-X^2-r^2)^3}
={1 \over 2 \pi R} \, \frac{Z^2}{(T^2-X^2-Z^2)^2}
\end{equation}
as promised.

Now let's return to the question of dealing with light-cone singularities in the CFT.
That is, let's ask how we can analytically continue this result outside the range (\ref{PoincareRange}).
In general the integrand in (\ref{PoincareContour}) has two double poles, located at
\begin{equation}
z = z_1 = - \frac{r}{T-X} \quad {\rm and} \quad z = z_2 = - \frac{T+X}{r}\,.
\end{equation}
In the range (\ref{PoincareRange}) we see that the contour always encircles the pole at
$z_1$ and never encircles the pole at $z_2$.  When we try to go outside this range one of
the poles crosses the integration contour $\vert z \vert = 1$.  So to analytically continue
the calculation outside the range (\ref{PoincareRange}) we merely have to deform the $z$
contour of integration so that it continues to encircle the pole at $z_1$ and exclude the pole at $z_2$.

One might ask how one can distinguish the two poles in general. Recall that the boundary CFT correlator 
is defined with a $T \rightarrow T - i \epsilon$ prescription. This means the poles are displaced
to\footnote{Assuming that $T$ and $X$ are real and $r > 0$.}
\begin{equation}
z_1 = - \frac{r}{T-X} - i \epsilon \qquad z_2 = - \frac{T+X}{r} + i \epsilon \,.
\end{equation}
We see that $z_1$ is always in the lower half plane while $z_2$ is always in the upper half plane.  So the general
prescription is to only encircle the pole in the lower half plane.  The $i \epsilon$ prescription
makes the $z$ contour integral well-defined, since the poles never collide.  It also makes the integral over $r$
well-defined, since the poles in $r$ are displaced off the real axis.

This lets us see how the bulk light-cone singularity emerges from the CFT.  Let's perform the $z$ integral in
(\ref{PoincareContour}) first, followed by the $r$ integral.  The two poles pinch the $z$ contour of integration
when $r^2 = r_0^2 \equiv (T - i \epsilon)^2 - X^2$.  Thus the integral over $z$ has a pole when $r = \pm r_0$.
When one of these singularities in the complex $r$ plane hits the $r = Z$ endpoint of the contour for integrating
over $r$, the integral over $r$ diverges.  This reproduces the bulk light-cone singularity at $T^2 - X^2 = Z^2$,
regulated by the appropriate $i\epsilon$ prescription.

Since our smeared operators have the correct 2-point functions, it follows that at infinite $N$ they
commute as operators in the CFT whenever the bulk points are spacelike separated.  This relies on the fact
that at infinite $N$ the commutator is a c-number, and one can check that it vanishes at bulk spacelike separation
by computing a correlator $\langle \psi \vert {\cal O}_1 {\cal O}_2 - {\cal O}_2 {\cal O}_1 \vert \psi \rangle$
in any state of the CFT.  However at finite $N$ the commutator becomes an operator.  The delicate cancellations
which occurred at infinite $N$ become state-dependent and are no longer possible in general.  Thus we do not necessarily
expect the commutator to vanish at bulk spacelike separation.  We discuss this point further in section \ref{locality}.

\section{AdS${}_3$ in Rindler coordinates}\label{AdS3}

We now specialize to AdS${}_3$.  This is a particularly interesting example, since the BTZ black
hole can be constructed as a quotient of AdS${}_3$ \cite{Banados:1992wn}.  After some
preliminaries we discuss AdS${}_3$ in accelerating Rindler-like coordinates.  We show
that our Poincar\'e results can be translated into accelerating coordinates and, with the help of an
antipodal map, can be used to describe local bulk operators inside the Rindler horizon.

\subsection{Preliminaries}\label{AdS3prelims}

AdS${}_3$ can be realized as the universal cover of a hyperboloid
\be
-U^2 - V^2 + X^2 + Y^2 = -R^2
\ee
inside ${\mathbb R}^{2,2}$ with metric $ds^2 = - dU^2 - dV^2 + dX^2 + dY^2$.
\begin{figure}
\centerline{\includegraphics{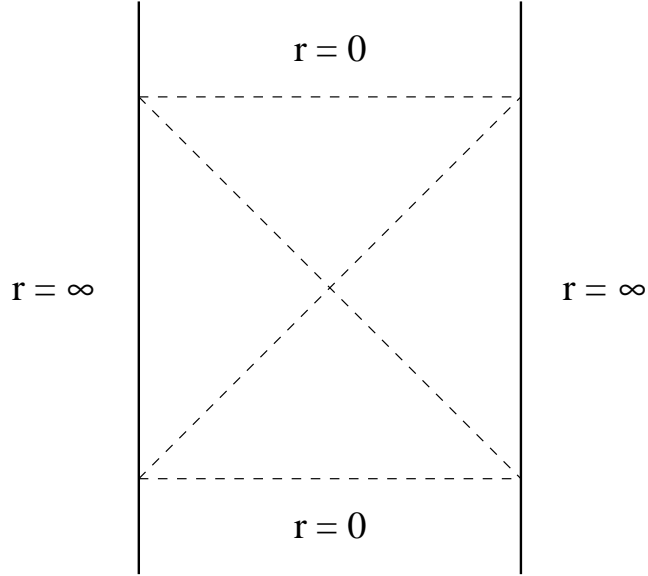}}
\caption{A slice of constant $\phi$ in AdS${}_3$, drawn as an AdS${}_2$ Penrose diagram.
The four Rindler wedges are separated by horizons at $r = r_+$.}
\label{R1}
\end{figure}
To describe this in Rindler coordinates we set
\bea
\nonumber
&& U = {R r \over r_+} \cosh {r_+ \phi \over R} \\
\label{AdS3embedding}
&& V = R \sqrt{{r^2 \over r_+^2} - 1} \sinh {r_+ t \over R^2} \\
\nonumber
&& X = R \sqrt{{r^2 \over r_+^2} - 1} \cosh {r_+ t \over R^2} \\
\nonumber
&& Y = {R r \over r_+} \sinh {r_+ \phi \over R}
\eea
so that the induced metric is
\be
\label{RindlerMetric}
ds^2 = - {r^2 - r_+^2 \over R^2} dt^2 + {R^2 \over r^2 - r_+^2} dr^2 + r^2 d\phi^2\,.
\ee
Here $-\infty < t,\phi < \infty$ and $r_+ < r < \infty$.  The Rindler horizon is located
at $r = r_+$.  These coordinates cover the right Rindler wedge of AdS${}_3$ as shown in
Fig.~\ref{R1}.  One can continue into the future wedge by setting
\beas
&& U = {R r \over r_+} \cosh {r_+ \phi \over R} \\
&& V = R \sqrt{1 - {r^2 \over r_+^2}} \cosh {r_+ t \over R^2} \\
&& X = R \sqrt{1 - {r^2 \over r_+^2}} \sinh {r_+ t \over R^2} \\
&& Y = {R r \over r_+} \sinh {r_+ \phi \over R^2}
\eeas
with $0 < r < r_+$.  One can extend these coordinates to the (left, past) wedges by
starting from the (right, future) definitions and changing the signs of $V$ and $X$.
It will frequently be convenient to work with rescaled coordinates
\[
\htt = r_+ t / R^2 \qquad \qquad \hp = r_+ \phi / R\,.
\]
An AdS-invariant distance function is provided by
\be
\label{AdSDistance}
\sigma(x \vert x') = - {1 \over R^2} X_\mu X'{}^\mu
\ee
in terms of the embedding coordinates $X^\mu = X^\mu(x)$.
For two points in the right Rindler wedge we have
\be
\label{sigmarindler}
\sigma = {r r' \over r_+^2} \cosh(\hp - \hp') - \left({r^2 \over r_+^2} - 1\right)^{1/2}
\left({r'^2 \over r_+^2} - 1\right)^{1/2} \cosh(\htt - \htt')
\ee
while for a point $(\htt,r,\hp)$ inside the future horizon and a point $(\htt',r',\hp')$ in the
$\left\lbrace{{\rm right} \atop {\rm left}}\right\rbrace$ Rindler wedge we have
\be
\label{FRAdSDistance}
\sigma = {r r' \over r_+^2} \cosh(\hp - \hp') \mp \left(1 - {r^2 \over r_+^2}\right)^{1/2}
\left({r'^2 \over r_+^2} - 1\right)^{1/2} \sinh(\htt - \htt')
\ee
%

\subsection{Rindler smearing functions}\label{Rindler}

We could set about constructing a smearing function starting
from a Rindler mode sum.  For points outside the Rindler horizon
this was carried out in \cite{hkll2}, while for points inside
the horizon we set up but do not evaluate the mode sum in
appendix \ref{RindlerModeSum}.  However the Rindler mode
sum is divergent and must be defined by analytic continuation in $\htt$ and/or $\hp$.
The divergence means there is no smearing function with support on real values of the
Rindler boundary coordinates.

\begin{figure}
\centerline{\includegraphics{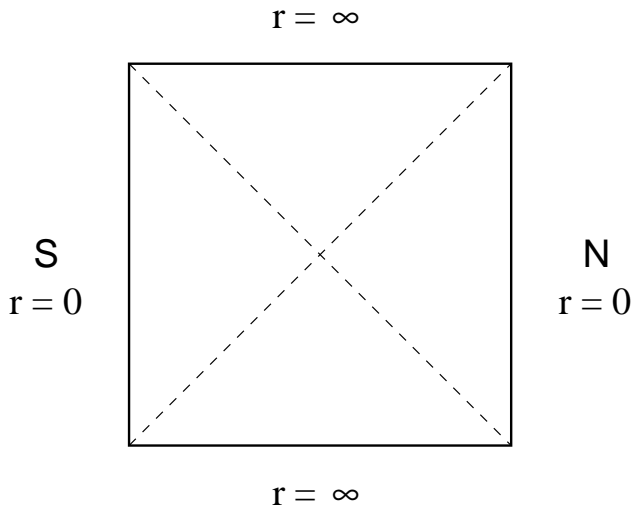}}
\caption{de Sitter space in static coordinates.}
\label{R2}
\end{figure}

A simpler approach to constructing the Rindler smearing function is to begin
with our Poincar\'e result (\ref{PoincareModeResult}) and translate it into
Rindler coordinates.  The translation is easiest to understand in de Sitter space.  Wick rotating
$\hp = i y$ turns the AdS metric (\ref{RindlerMetric}) into
\[
ds^2 = {R^2 \over r_+^2} \left[{r_+^2 \over r^2 - r_+^2} dr^2
- (r^2 - r_+^2) d\htt^2 - r^2 dy^2\right]\,.
\]
This is de Sitter space in static coordinates.  To avoid a conical
singularity at $r = 0$ we must periodically identify $y \sim y + 2\pi$.
The right Rindler wedge becomes the past wedge of de Sitter space, as shown
in Fig.~\ref{R2}.  The induced metric on the past boundary is,
up to a divergent conformal factor,
\[
ds^2_{\rm bdy} = d\htt^2 + dy^2 \qquad -\infty < \htt < \infty\,, \quad y \sim y + 2 \pi
\]
i.e.\ an infinite cylinder which can be compactified to a sphere by adding the
north and south poles.  This sphere
can be identified with the past boundary that we identified working in Poincar\'e
coordinates.  However note that any observer inside the past wedge of de Sitter
space can at most see one hemisphere of the past boundary, namely the region
characterized by
\[
-\infty < \htt < \infty \qquad -\pi/2 < y < \pi/2\,.
\]

For a point inside the past wedge of de Sitter we can construct a retarded Green's
function that lets us express the value of the field in terms of data on the past
boundary.  In AdS this means we can express the value of the field anywhere in the
right Rindler wedge in terms of a data on the right Rindler boundary.  In fact the
result is a simple translation of our Poincar\'e result (\ref{PoincareModeResult}).
We define the right boundary field in Rindler coordinates by
\be
\phinotrR{}(\htt,\hp) = \left. \lim_{r \rightarrow \infty} r^\Delta \phi(\htt,r,\hp)
\right\vert_{\hbox{\small right boundary}}
\ee
This is related to the Poincar\'e boundary field by
\be
\phinotrR{}(\htt,\hp) = \lim_{r \rightarrow \infty} (rZ)^\Delta \phinotp{}(T,X)\,.
\ee
We also have the boundary change of coordinates
\be
{dT dX \over Z^2} = {r^2 d\htt d\hp \over r_+^2}\,.
\ee
Making these substitutions in (\ref{PoincareModeResult}), the value of the field at a
bulk point inside the right Rindler wedge of AdS${}_3$ is
\be
\label{RightRindler}
\phi(\htt,r,\hp) = {(\Delta - 1) 2^{\Delta - 2} \over \pi r_+^2} \int_{\hbox{\small spacelike}}
\hspace{-0.5cm} dx dy \,
\lim_{r' \rightarrow \infty} (\sigma / r')^{\Delta - 2} \phinotrR{}(\htt + x, \hp + i y)
\ee
where as $r' \rightarrow \infty$ the AdS invariant distance (\ref{sigmarindler}) becomes
\be
\sigma(\htt,r,\hp \vert \htt + x,r',\hp + iy) = {rr' \over r_+^2} \left[\cos y -
\left(1 - {r_+^2 \over r^2}\right)^{1/2} \cosh x \right]
\ee
and the integration is over ``spacelike separated'' points on the Wick rotated boundary,
that is, over real values of $(x,y)$ such that $\sigma > 0$.

The result (\ref{RightRindler}) for bulk points in the right Rindler wedge was obtained in \cite{hkll2},
starting from a Rindler mode sum and defining it via an analytic continuation,
or alternatively from a de Sitter Green's function. Now let's ask what happens for
bulk points inside the Rindler horizon.  It's clear from Fig.~\ref{R1} that, if we were willing to
work in Poincar\'e coordinates, there would be no problem: we could use (\ref{PoincareModeResult}) to obtain
a smearing function with compact support on the Poincar\'e boundary.  However if we wish to work in Rindler
coordinates there is a problem: the smearing function extends outside the Rindler wedge, and covers points
on the (real slice of) the boundary which are to the future of the right Rindler patch.\footnote{This
is easiest to see by considering a bulk point in the future wedge of Fig.~\ref{R1} and
following light rays to the right boundary.}

To fix this we can use the antipodal map.\footnote{Exactly the same procedure applies to AdS${}_2$
in Rindler coordinates \cite{hkll}.  An alternate procedure would be to analytically continue outside the
strip $-\pi/2 < y < \pi/2$.}  The antipodal map acts on the embedding coordinates of section
\ref{AdS3prelims} by
\be
A \, : \, X^\mu \rightarrow - X^\mu\,.
\ee
In terms of Rindler coordinates this can be realized by
\be
A \, : \, \htt \rightarrow \htt + i \pi,\quad \hp \rightarrow \hp + i \pi\,.
\ee
Note that $\sigma(x \vert Ax') = - \sigma(x \vert x')$.
Fields with integer conformal dimension transform simply under the
antipodal map,
\be
\phi(Ax) = (-1)^\Delta \phi(x)\,.
\ee
This is discussed in appendix \ref{NonIntegerDelta}, where we also treat the slightly more
involved case of non-integer $\Delta$.

In Rindler coordinates the antipodal map can be used to move the part of the
smearing function which extends outside the right Rindler wedge over to the left
boundary.  To see this one starts with the Poincar\'e result (\ref{PoincareModeResult}) and
breaks the integration region up into two pieces.  One piece gives a smearing function in
the right Rindler wedge, while under the antipodal map the other piece becomes a smearing
function in the left Rindler wedge.  Thus for a bulk point inside the Rindler horizon we have
\bea
\nonumber
\phi(\htt,r,\hp) & = & {(\Delta - 1) 2^{\Delta - 2} \over \pi r_+^2} \biggl[\int_{\sigma > 0}
\hspace{-0.5cm} dx dy \,
\lim_{r' \rightarrow \infty} (\sigma / r')^{\Delta - 2} \phinotrR{}(\htt + x, \hp + i y) \\
\label{InsideRindler}
& & \!\!\!\!\!\! + \int_{\sigma < 0}
\hspace{-0.5cm} dx dy \,
\lim_{r' \rightarrow \infty} (-\sigma / r')^{\Delta - 2} (-1)^\Delta \phinotrL{}(\htt + x, \hp + i y) \biggr]\,.
\eea
Here as $r' \rightarrow \infty$ the AdS invariant distance (\ref{AdSDistance}) becomes
\be
\sigma(\htt,r,\hp \vert \htt + x,r',\hp + iy) = {rr' \over r_+^2} \left[\cos y \mp \left({r_+^2 \over r^2} - 1\right)^{1/2}
\sinh x \right]
\ee
when the boundary point is in the $\left\lbrace{{\rm right} \atop {\rm left}}\right\rbrace$ Rindler wedge.
The integration is over points with $\sigma > 0$ on the right boundary and points with $\sigma < 0$ on the left boundary,
and we define
\be
\phinotrL{}(\htt,\hp) = \left. \lim_{r \rightarrow \infty} r^\Delta \phi(\htt,r,\hp)
\right\vert_{\hbox{\small left boundary}}\,.
\ee
%

\subsection{Reproducing bulk correlators}\label{correlators2}

It is instructive to check that the Rindler smearing functions we have constructed let us recover
the correct bulk two-point functions from the CFT,\footnote{This was done in section \ref{correlators} for Poincar\'e coordinates.}
especially for points inside the Rindler horizon.
Clearly of special importance is the point $r=0$, where the Rindler
coordinates become singular.  So in this section we show how this works for a point located at $r=0$
and a point near the right boundary.

The AdS Wightman function is 
\be
\label{AdSWightman}
G_{\rm AdS}(x \vert x') = \langle 0 \vert \phi(x) \phi(x') \vert 0 \rangle_{SUGRA} = {1 \over 4 \pi R}
{1 \over \sqrt{\sigma^2 - 1}} {1 \over (\sigma + \sqrt{\sigma^2 - 1})^{\Delta-1}}\,.
\ee
Here $\vert 0 \rangle$ is the global or AdS-invariant vacuum state.  Branch cuts are handled with
a $\tau \rightarrow \tau - i \epsilon$ prescription, or equivalently $\sigma \rightarrow \sigma + i \epsilon \sin (\tau - \tau')$,
where $\tau$ is the global time coordinate defined in appendix \ref{NonIntegerDelta}.\footnote{For points inside the
Poincar\'e patch this is equivalent to $T \rightarrow T - i \epsilon$.}
We consider a point near the origin of Rindler coordinates  $(t = 0, r = r_0, \phi = 0)$, and a
point near the right boundary with coordinates $(t,r,\phi)$.  As $r_0 \rightarrow 0$ and $r \rightarrow \infty$ the
invariant distance (\ref{FRAdSDistance}) is
\[
\sigma \approx {r \over r_+} \left({r_0 \over r_+} \cosh \hp + \sinh \htt \right)\,.
\]
Thus the AdS correlator approaches a finite, $\hp$-independent value as $r_0 \rightarrow 0$
\be
\label{AdSCenterCorrelator}
G_{\rm AdS}(0,0,0 \vert \htt,r,\hp) \approx {1 \over 2 \pi R}
\left({r_+ \over 2 r \sinh \htt + i \epsilon}\right)^\Delta~.
\ee
The fact that the correlator is independent of $\hp$ reflects the fact that $r = 0$ is a fixed point
of the isometry $\hp \rightarrow \hp + {\rm const}$.

Now let's see how this behavior is reproduced by the CFT.  We'll work with a field of integer
conformal dimension.   At $\htt = r = \hp = 0$ the smearing function (\ref{InsideRindler})
reduces to
\be
\label{CenterSmear}
\phi(0,0,0) = {(\Delta - 1) 2^{\Delta - 2} \over \pi r_+^\Delta}
\int_0^\infty dx \sinh^{\Delta-2}x \int_{-\pi/2}^{\pi/2} dy \left(\phinotrR{}(x,iy) + (-1)^\Delta
\phinotrL{}(x,iy)\right)
\ee
while as $r \rightarrow \infty$ the smearing function (\ref{RightRindler}) reduces to
\be
\phi(\htt,r,\hp) \approx r^{-\Delta} \phinotrR{}(\htt,\hp)\,.
\ee
This means we should be able to recover (\ref{AdSCenterCorrelator}) by computing
\bea
\label{AdScomputation}
& & {(\Delta - 1) 2^{\Delta - 2} \over \pi (r r_+)^\Delta}
\int_0^\infty dx \sinh^{\Delta-2}x \int_{-\pi/2}^{\pi/2} dy \\
\nonumber
& & \qquad\qquad\qquad \Big\langle \Big(\phinotrR{}(x,iy) + (-1)^\Delta \phinotrL{}(x,iy)\Big)
\phinotrR{}(\htt,\hp) \Big\rangle_{CFT}~.
\eea
For convenience we'll work in the regime
\be
\label{EasyRegime}
\quad \htt < 0, \quad \htt < \hp < -\htt\,.
\ee
In this regime the smeared CFT operators are never lightlike separated, so (\ref{AdScomputation}) is well
defined without a prescription for dealing with lightcone singularities in the CFT.
The appropriate CFT correlators can be obtained from (\ref{AdSWightman}) by
sending the bulk points to the appropriate boundary\footnote{We obtained these
expressions as boundary supergravity correlators, but the same result holds for correlation functions
in a finite temperature CFT.}
\bea
\nonumber
&&\hspace{-1.5cm}\langle \phinotrR{}(\htt,\hp) \phinotrR{}(\htt',\hp') \rangle_{CFT} = {(r_+^2/2)^\Delta \over
2 \pi R \left(\cosh(\hp - \hp') - \cosh(\htt - \htt' - i \epsilon)\right)^\Delta} \\
\nonumber
&&\hspace{-1.5cm}\langle \phinotrR{}(\htt,\hp) \phinotrL{}(\htt',\hp') \rangle_{CFT} = {(r_+^2/2)^\Delta \over
2 \pi R \left(\cosh(\hp - \hp') + \cosh(\htt - \htt')\right)^\Delta} ~.\\
\label{AdSBdyCorr}
\eea
Then we have
\bea
\nonumber
& & {(\Delta - 1) r_+^\Delta \over 8 \pi^2 R r^\Delta} \int_0^\infty dx \sinh^{\Delta - 2}x
\int_{-\pi/2}^{\pi/2} dy \,
\Biggl[\left(\cosh(\hp - iy) - \cosh(\htt - x)\right)^{-\Delta} \\
\nonumber
& & \hspace{4cm} + (-1)^\Delta \left(\cosh(\hp - iy) + \cosh(\htt - x)\right)^{-\Delta}\Biggr] \\
\nonumber
& = & {(\Delta - 1) r_+^\Delta \over 8 \pi^2 R r^\Delta} \int_0^\infty dx \sinh^{\Delta - 2}x \int_{-\pi}^\pi dy \,
\left(\cosh(\hp - iy) - \cosh(\htt - x)\right)^{-\Delta} \\
\label{ContourIntegral}
& = & {(\Delta - 1) 2^{\Delta - 3} r_+^\Delta \over i \pi^2 R r^\Delta} \int_0^\infty dx \sinh^{\Delta - 2}x
\oint_{\vert z \vert = e^{\phi}} {z^{\Delta - 1} dz \over (z - e^{\htt - x})^\Delta (z - e^{-(\htt - x)})^\Delta}
\eea
In the last line we set $z = e^{\hp - i y}$.  In the regime (\ref{EasyRegime}) the $z$ contour of integration
always encircles the pole at $e^{\htt - x}$ and never encircles the pole at $e^{-(\htt - x)}$.  To analytically
continue outside (\ref{EasyRegime}) we proceed as in section \ref{correlators} and deform the contour of integration so that it
continues to encircle the appropriate pole.  This continuation gives an integral that
is independent of $\hp$, and in this way the smearing function (\ref{CenterSmear}) captures the fact that $r = 0$
is a fixed point of the isometry $\phi \rightarrow \phi + {\rm const}$.  
It's entertaining to push the calculation a little further and show that the CFT exactly reproduces
the bulk correlator.  Just to be concrete, let's set $\Delta = 2$.  Then evaluating
the contour integral in (\ref{ContourIntegral}) gives
\be
-{r_+^2 \over \pi R r^2} \int_0^\infty dx {2 \cosh(\htt - x) \over \left(2 \sinh(\htt - x)\right)^3}
=  {1 \over 2 \pi R} \left({r_+ \over 2 r \sinh \htt}\right)^2
\ee
in agreement with (\ref{AdSCenterCorrelator}) for $\Delta = 2$. The result is also valid outside the range (\ref{EasyRegime})
using the analytic continuation described above.

\section{BTZ black hole}\label{BTZ}

To make a BTZ black hole starting from AdS${}_3$ all we have to do is periodically identify the
$\phi$ coordinate, $\phi \sim \phi + 2\pi$ \cite{Banados:1992wn,Carlip:1995qv}.  Scalar fields on
AdS${}_3$ will descend to scalar fields on BTZ provided they satisfy $\phi(t,r,\phi) = \phi(t,r,\phi + 2\pi)$.
The global AdS vacuum descends to the Hartle-Hawking vacuum state in BTZ.

In this construction we are identifying points separated by {\em real}
values of the $\phi$ coordinate.  Since the Rindler smearing functions we have constructed
are translation invariant in $\phi$, and since they only involve integration over the imaginary part of
$\phi$, we can apply our Rindler results to a BTZ black hole without modification.
That is, (\ref{RightRindler}) and (\ref{InsideRindler}) can be used to represent bulk fields in a BTZ
spacetime; if the boundary field has the correct periodicity then so will the bulk field.
This shows quite explicitly that we can recover local physics outside the BTZ horizon using
operators that act on a single copy of the CFT, while to describe the region inside the
horizon we must use operators that act on both the CFT and its thermofield double
\cite{hkll,Kraus:2002iv,Fidkowski:2003nf,Festuccia:2005pi,Maldacena:2001kr}.

The BTZ black hole has a spacelike singularity at $r = 0$, which has been studied from the CFT
point of view in \cite{Fidkowski:2003nf,Festuccia:2005pi,Hubeny:2006yu,Yang:2006pc}.\footnote{The singularity is an analytic
continuation of one of conical type.  The curvature remains constant near the singularity.}
In the semiclassical limit that we are considering this singularity should be encoded in the CFT.
We are now in a position to see this directly, by studying bulk correlators with one point close to
the singularity.

The BTZ Wightman function is given by an image sum \cite{Lifschytz:1993eb,Ichinose:1994rg}
\be
\label{gbtz}
G_{\rm BTZ}(x \vert x') = \sum_{n = -\infty}^\infty G_{\rm AdS}(t,r,\phi \vert t',r',\phi' + 2 \pi n)~.
\ee
This diverges when $r = 0$, just because $r = 0$ is a fixed point of the isometry of shifting $\phi$ by a constant: when $r = 0$
the invariant distance (\ref{FRAdSDistance}) is independent of $\phi'$
and the image sum diverges.  To estimate the divergence, note that for small $r_0$ the BTZ image sum
is cut off at $\vert n \vert \approx {1 \over 2 \pi} \log (r_+/r_0)$.  This
means the BTZ Wightman function diverges logarithmically near the singularity
\[
G_{\rm BTZ}(0,r_0,0 \vert \htt,r,\hp) \sim {1 \over 2 \pi^2 R} \left({r_+ \over 2 r \sinh \htt}\right)^\Delta
\log {r_+ \over r_0} \quad \hbox{\rm as $r_0 \rightarrow 0$}~.
\]

How does this divergence arise from the CFT viewpoint?  A priori there are a number of possibilities:
\begin{itemize}
 \item The CFT itself could be incomplete in the same sense as classical gravity.
 \item The mapping between CFT operators and local bulk fields could become singular at this point.
 \item The mapping could remain smooth, but the CFT operator moves outside the class of physically reasonable observables.
\end{itemize}
The boundary S-matrix in the gravity theory appears to be well-defined around the BTZ background by virtue of cosmic censorship, provided one avoids processes that produce naked singularities. Hence the same will be true of the CFT correlators, so in that sense the CFT gives a complete well-defined theory at large $N$.
 Thus the first possibility is ruled out. The mapping is non-singular, as can be seen explicitly
in (\ref{CenterSmear}), which rules out the second possibility.  It is the third possibility which is realized.

Before discussing this in more detail, let us follow through with our calculation of the bulk two point function using the CFT.
As in section \ref{correlators2} we place one point near the singularity and the other near the right boundary.  Then all we have
to do is replace the AdS boundary correlators with BTZ boundary
correlators in (\ref{AdScomputation}).\footnote{As written, (\ref{AdScomputation}) is only valid in the range (\ref{EasyRegime}).
To extend it outside this range we must analytically continue in $\phi$, as discussed at the end of section \ref{correlators2}.}  Boundary
correlators in the BTZ geometry can be obtained from (\ref{AdSBdyCorr}) by performing an image sum to make them $2\pi$ periodic in $\phi$
\cite{Keski-Vakkuri:1998nw}.  However as we have seen (\ref{AdScomputation}) gives a result that is independent of $\phi$.
Therefore substituting BTZ boundary correlators in (\ref{AdScomputation}) leads to a divergent image sum.  So the divergence
is present in the CFT computation of the correlator, for  the same reason it was present in the bulk.

Now let's make some comments on the interpretation of this divergence.  In AdS${}_3$ as two bulk points coincide their correlator
exhibits the expected Hadamard short-distance singularity
\be
\label{hadamard}
 G_{\rm AdS} \sim \frac{1}{4\pi R\sqrt{2(\sigma-1)}} \quad \hbox{\rm as $\sigma \rightarrow 1$}~.
\ee
Generically as two points coincide in BTZ their correlator diverges in exactly the same way, because only one term in the
image sum (\ref{gbtz}) will have a singularity.  However if we place one point at the BTZ singularity then $G_{\rm BTZ}$
diverges no matter where the other point is located.  This is because $r=0$ is a fixed point of the orbifold symmetry
and the symmetry operation is of infinite order.

We can use the coefficient of the singularity (\ref{hadamard}) as a definition of the norm of these operators. For generic points
the norm is finite, however the norm diverges for the operator at the fixed point. One way to see this is by using a point splitting
regularization and considering $\lim_{\epsilon\to 0} G_{\rm BTZ}(0,0,0 \vert \epsilon_r,\epsilon_\phi,\epsilon_t)$.  The invariant
distance is independent of the coordinate separation in the $\phi$ direction if one point lies at $r=0$, so
$G_{\rm BTZ}(0,0,0\vert \epsilon_r,\epsilon_\phi,\epsilon_t)$ diverges even at finite $\epsilon$.  Thus the operator $\phi\vert_{r=0}$
has infinite norm.

In the CFT we interpret the operator (\ref{CenterSmear}) dual to $\phi\vert_{r=0}$ exactly as in the bulk. It is a non-normalizable operator which
has divergent correlators with all operators of interest.  This is how a well-behaved conformal field theory gives rise to
a divergent correlation function: through the introduction of a non-normalizable operator. We will comment further in section \ref{conclusions} on how this picture generalizes when back-reaction and finite $N$ are taken into account.

\section{Collapse geometries}\label{collapse}

As we have seen, it is possible to probe the region inside the horizon of a BTZ black hole
using operators that act on both the left and right copies of the CFT.  A similar
result should hold for a general eternal AdS-Schwarzschild black hole.  However in the more
physical case of a black hole formed in collapse there is only a single asymptotic AdS region, and
one might ask: can the region inside the horizon be described using the single copy of the CFT?

\begin{figure}
\centerline{\includegraphics{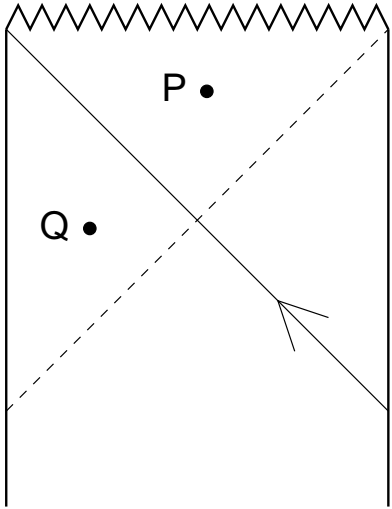}}
\caption{An AdS black hole formed by collapse.  The left edge of the diagram is the origin of AdS,
the right edge is the AdS boundary.  The dashed line is the black hole horizon while the solid diagonal line
represents the infalling shell.}
\label{C}
\end{figure}

For simplicity let's work in AdS${}_3$ and consider a large (stable) black hole formed by sending in a null shell from the
boundary.  The Penrose diagram is shown in Fig.~\ref{C}.  Consider a bulk point {\cmss P}
inside the horizon and to the future of the shell.  Can an operator inserted at that point be described in the CFT?

The answer is yes, and for fields with integer conformal dimension the explicit construction is quite simple.
As can be seen from the global mode expansion given in appendix \ref{NonIntegerDelta}, fields with integer
conformal dimension are single-valued on the AdS hyperboloid (periodic in global time with period $2\pi$).
Note from (\ref{AdS3embedding}) that continuing $\htt \rightarrow \htt + i \pi$ has the effect of changing
the sign of two of the embedding coordinates, namely
\[
V \rightarrow -V \qquad X \rightarrow -X\,.
\]
Thus for integer conformal dimension the boundary fields in the left and right Rindler wedges are related by
\be
\label{LeftRightRelation}
\phinotrL{}(\htt,\hp) = \phinotrR{}(\htt + i \pi,\hp)\,.
\ee
(This relation was also used in \cite{Kraus:2002iv}).
The collapse geometry can be made by taking the right and future regions of an eternal BTZ black hole and
joining them across the shell to a piece of AdS${}_3$.  For points to the future of the shell we can, by
analytic continuation, pretend that we are in an eternal BTZ geometry.  We can therefore use the relation
(\ref{LeftRightRelation}) in our BTZ smearing function (\ref{InsideRindler}) to represent bulk operators
that are located inside the horizon.

This shows that we can represent a bulk point inside the horizon in terms of a single CFT, provided
we analytically continue in both the $\htt$ and $\hp$ coordinates.  Our explicit construction works
for points that are to the future of the infalling shell.  One could also ask about representing bulk points
inside the shell, such as the point {\cmss Q} in Fig.~\ref{C}.  This is indeed possible, although the construction is
more complicated since one must propagate modes across the shell \cite{Roy}.

\section{Comments on finite $N$}\label{locality}

We have seen that in the semiclassical limit one can construct local
operators anywhere in the bulk of AdS.  However at finite $N$, when the Planck
length is finite, holography demands that the number of independent degrees of freedom inside a volume is finite, bounded by the area of the region
in Planck units.  In this section we attempt to understand how this comes about.

The smeared operators we have constructed in the CFT are still well-defined at finite $N$.
For example in ${\cal N} = 4$ Yang-Mills
we can define the operator
\be
\label{FiniteN}
\Phi(T,X,Z) = \int dT' d^3X' \, K(T',X' \vert T,X,Z) \, {\rm Tr} \, F^2(T',X')
\ee
at any $N$.  At
finite $N$ it does not obey the correct bulk dilaton equation of
motion \cite{Banks:1998dd}. However it is a
perfectly good operator in the gauge theory, and it has the right
behavior in the large-$N$ limit to be associated with a particular
point in the bulk.  So as a first step,  it seems reasonable to associate $\Phi(T,X,Z)$
with a point in the bulk, even at finite $N$.  Since the bulk point
was arbitrary, at first sight this means we can associate an infinite number of local
operators with any given region in the bulk.

This might seem like a surprising conclusion, so let us give supporting evidence for our approach.
Consider pure AdS${}_D$, dual to a CFT${}_{D-1}$ in its ground state.  The conformal symmetry
of the CFT is valid at any $N$.  This means that, even when the
Planck length is finite, AdS quantum gravity has an exact $SO(D-1,2)$ symmetry.  Purely formally, we can realize
this symmetry as acting on a set of coordinates $(T,X,Z)$.
The smearing functions we have constructed transform covariantly under $SO(D-1,2)$ \cite{hkll2} -- a property which suffices
to determine them up to an overall coefficient.\footnote{The coefficient can be
fixed by matching onto a properly normalized operator in the CFT in the limit that the bulk point approaches the boundary.}
This means that at any $N$, the smearing functions we have defined are the unique way to start with a primary operator
in the CFT and build a representation of $SO(D-1,2)$ which transforms as a scalar field in AdS.  Since the construction we
have outlined is fixed by the symmetries, the operators (\ref{FiniteN}) are singled out even at finite $N$.

How can this continuum of operators be compatible with holography?
We believe the answer is that only a few of these operators will commute with each other at finite $N$. At infinite $N$ we managed to construct smeared operators in the CFT which commute with each other
even though the smearing functions overlap on the boundary.  We discussed this
in section \ref{correlators}.  But at finite $N$ it is implausible
that all the overlapping operators will commute.\footnote{By overlapping operators we mean the smearing functions
have support at timelike relative separation.}  Let's estimate how many commuting operators we do expect.
If we take a local CFT operator and smear it, it will trivially
commute with another smeared operator provided the two smearings are
``spacelike'' to each other: that is, provided the two smearing
functions have supports which only involve points on the boundary that
are at spacelike separation.  In this case the two smeared operators
will commute with each other by locality of the boundary theory.
The condition for spacelike separation was studied in \cite{hkll} for
AdS${}_3$ and is easily extended to any dimension.  In Poincar\'e
coordinates, working on a hypersurface of fixed time, it boils down to
the requirement that the separation between any two bulk
operators satisfies $\vert \Delta X \vert > 2Z$.  Since the necessary
separation gets larger as $Z$ increases, the maximum number of
commuting operators in a given region is achieved by placing them all
at the boundary of the region.  For example, inside a bulk region
\[
0 < X_{i} < L \qquad Z_0 < Z < \infty \qquad\quad i=1 \cdots d-1
\]
the maximum number of trivially commuting operators is given when they are
evenly spaced along the boundary of the region, at $Z = Z_0$, with a
characteristic coordinate spacing of order $Z_0$.  Thus according to
this prescription there are at most $\sim (L/Z_0)^{d-1}$ trivially commuting
operators one can build in this region by smearing a single local
operator in the CFT \footnote{Thus the bulk
region can be described by a boundary theory with a lattice spacing
$\sim Z_0$.  This is clearly closely related to the cutoff procedure
introduced in \cite{Susskind:1998dq}.}. This corresponds to one commuting
operator per AdS area (in units of the AdS radius of curvature $R$). This is far too few
degrees of freedom to describe a local bulk field.

Turning back to the infinite number of operators described above, we note that -- although they do not all commute --
their correlation functions nonetheless look local up to $1/N$ corrections that involve mixing with other operators.  The infinite set of operators can therefore be used to describe bulk
physics which is approximately local, at least as far as correlation functions are concerned, as long as the $1/N$ corrections can
be ignored.  However note that if one tries to place operators too close together or in a state with large energy,  their commutator may get
a large contribution from smeared operators corresponding to bulk excitations which are outside the given spacetime volume. We should not associate such operators with independent degrees of freedom within the volume. Presumably there is a finite maximal set of operators that mutually commute up to terms that vanish as $N\to\infty$ and remain inside the given volume. Bekenstein-style arguments \cite{Bekenstein:1980jp} (made on the supergravity side) support this idea.
It is this set of operators which we argue counts the independent degrees of freedom inside a volume.

We obtain a natural proposal for a basis of these operators by generalizing the above construction of trivially commuting operators.
Let us consider all possible degrees of freedom within a given bulk volume, rather than those associated with a particular supergravity field. For concreteness, we will consider AdS${}_5$.
We expect of order $N^2$ independent local operators in the
boundary theory. Therefore we should be able to construct a basis of $N^2$ mutually commuting bulk operators as we did above for the trivially commuting operators. This implies a total of $N^2$ degrees of freedom per
area in AdS units.  This matches perfectly with the relation $l_{\rm
Planck}^3 = R^3/N^2$ and saturates the holographic bound. 

\section{Conclusions}\label{conclusions}

In this paper we developed the representation of local operators in
the bulk of AdS in terms of non-local operators on the complexified
boundary.  We showed that these non-local operators reproduce the
correct bulk-to-bulk correlation functions.  In particular they
reproduce the divergent correlators of an operator inserted at the BTZ
singularity.  We commented on black holes formed by collapse, and
discussed the way in which bulk locality arises in the large-$N$
limit but breaks down at finite $N$.  

Local bulk operators thus provide a powerful tool for understanding
the AdS/CFT correspondence.  They give new insights into the way in
which light-cone singularities and spacelike commutativity arise in
the bulk.  They enable us to probe non-trivial geometries, including
regions inside horizons which are naively hidden from the boundary,
and they show very explicitly how a bulk singularity can manifest
itself in a well-behaved CFT.  Our results were all obtained in the
infinite $N$ limit.  However we argued that in some contexts (2-point
functions in pure AdS) our results carry over exactly to any value of
$N$.  And based on consideration of the operators at infinite $N$ we
were able to give a qualitative picture of the independent bulk degrees of
freedom at finite $N$.

There are a number of directions for future work. We begin with
a few further remarks on the nature of the BTZ singularity from the CFT viewpoint.
At leading order in a large $N$ expansion, we found that a bulk field probe of the singularity is represented by a non-normalizable operator in the CFT. Note that back-reaction/finite $N$ effects play a crucial role in understanding the physics near the singularity, even in the case of BTZ, as discussed in \cite{Lifschytz:1993eb,Berkooz:2002je,Kraus:2002iv} (and references therein). Therefore we certainly expect large corrections to the smearing function within a Planck length of the singularity. It would be interesting to know whether these corrections render operators at the singularity normalizable, or whether one should simply abandon a bulk spacetime description of the physics in this region. Nevertheless it seems the operators defined by  (\ref{InsideRindler}) have smooth analytic continuations through complex values of $r$ from region $2_{++}$ (in the notation of \cite{Kraus:2002iv}) to the past of the singularity to region $2_{-+}$ to the future of the singularity, avoiding the Planck scale region near the singularity. This raises the question of whether the CFT also gives a smooth description of regions to the future of the singularity.
An important criterion in deciding whether certain combinations of CFT correlators reproduce sensible bulk
spacetime physics, is whether the set of amplitudes can be reproduced by a unitary local bulk Lorentzian spacetime
effective action. This seems to be true for regions outside the horizon, and regions to the past of the
singularity, but it is unlikely this will be true if one also includes operators to the future of the singularity. It would be very
interesting show this explicitly. Moreover the resolution of the black hole information problem via AdS/CFT
suggests \cite{Lowe:1999pk,Lowe:2006xm} that non-local terms appear in a bulk effective action connecting the
region near the singularity with the region outside the horizon. The local operators constructed in the present
work are an important first step in trying to reconstruct these new quantum gravity features of the bulk effective
action. 

In section \ref{locality} we commented on the way in which the number
of commuting degrees of freedom is reduced at finite $N$.  We also
showed how bulk locality is recovered in correlation functions in the large-$N$ limit, despite
the seemingly low number of degrees of freedom (corresponding to a theory with a cutoff $\Delta X > Z$): namely, through the presence
of a continuum of bulk operators whose commutators are ${\cal
O}(1/N)$.  Constructing a precise analog of smearing functions at finite $N$ and better understanding the analog of bulk spacetime geometry is an important open problem.

For eternal black holes we found that local operators inside the
horizon are dual to operators which act on both copies of the CFT.  In
section \ref{collapse} we showed that, at least in some cases, one
could represent an operator inside the horizon of a black hole formed
by collapse in terms of a single CFT, by using an operator which is
analytically continued both in the spatial and temporal coordinates of
the CFT.  These ideas will be further explored in \cite{Roy}.

This leads to an interesting question, namely, whether there is
an algorithm for constructing smearing functions with compact support
in a general asymptotically AdS background.  The smearing functions we
have constructed in this paper can all be thought of as arising from a
Wick rotation of the boundary spatial coordinates. This should certainly be a well-defined operation on the analytic correlators that arise from the CFT.  However a general bulk geometry will typically not have an interpretation with a real metric after performing such a continuation.
 One could still try to
represent the smearing function as a mode sum, but it is not clear
that the smearing function will have compact support on the
(complexified) boundary.  One way to address this issue would be to
attempt to find a procedure, purely within the CFT, for identifying a
set of well-behaved smearing functions.  The only obvious condition to
impose is that in the semiclassical limit the smeared operators should
commute at bulk spacelike separation.  Is that enough to uniquely
determine the smearing functions?

\bigskip
\centerline{\bf Acknowledgements}
\noindent
We thank Raphael Bousso, Asad Naqvi and Shubho Roy for valuable discussions.
DK and DL are grateful to the 2006 Simons Workshop for hospitality.
The work of AH and DK is supported by US Department of Energy grant
DE-FG02-92ER40699. AH is supported in part by the Columbia University Initiatives in Science and Engineering. GL is supported in part by the Israeli science foundation,
grant number 568/05.  The research of DL is supported in part by DOE grant
DE-FG02-91ER40688-Task A.

\appendix
\section{Rindler mode sum}\label{RindlerModeSum}

In this appendix we set up the Rindler mode sum for a bulk point
inside the horizon.  It's convenient to introduce Kruskal coordinates on AdS${}_3$
in which
\[
ds^2 = - {4 R^2 \over (1 + uv)^2} du dv + r^2 d\phi^2\,.
\]
These coordinates are defined by
\beas
&& u = \left({r - r_+ \over r + r_+}\right)^{1/2} e^{\htt} \\
&& v = - \left({r - r_+ \over r + r_+}\right)^{1/2} e^{-\htt}
\eeas
in the right Rindler wedge and
\beas
&& u = \left({r_+ - r \over r_+ + r}\right)^{1/2} e^{\htt} \\
&& u = \left({r_+ - r \over r_+ + r}\right)^{1/2} e^{-\htt}
\eeas
in the future Rindler wedge; to cover the left and past wedges just
change the signs of both $u$ and $v$.  A complete set of normalizable
modes in the right Rindler wedge is given by
\be
\phi_R(t,r,\phi) = e^{-i \omega t} e^{i k \phi} r^{-\Delta} \left(1 - {r_+^2 \over r^2}\right)^{-i\hw/2}
F\left({\Delta \over 2} - i \hw^+,{\Delta \over 2} - i \hw^-,\Delta,{r_+^2 \over r^2}\right)
\ee
where $\omega,k \in {\mathbb R}$, $\hw^\pm = {1 \over 2} (\hw \pm \hk)$, $\hw = \omega R^2 / r_+$, $\hk = k R / r_+$.
We can extend this mode to the entire Kruskal diagram by analytically continuing across the Rindler horizons.
If we continue through the lower half of the complex $u$ and $v$ planes we get a mode which is positive frequency
with respect to Kruskal time, while continuing through the upper half of the complex $u$ and $v$ planes gives a
negative frequency Kruskal mode.\footnote{Positive and negative frequency in the sense of
multiplying annihilation and creation operators in the expansion of the field.  This prescription for selecting
positive frequency Kruskal modes picks out the AdS-invariant vacuum state.}  The analytic continuation is straightforward,
with the help of a $z \rightarrow 1 - z$ transformation of the hypergeometric function.  Define
\beas
&& f_{\omega k}(r) = {1 \over r^\Delta} \left(1 - {r_+^2 \over r^2}\right)^{-i \hw / 2} F\left({\Delta \over 2}
- i \hw^+, {\Delta \over 2} - i \hw^-, \Delta, {r_+^2 \over r^2}\right) \\
&& g_{\omega k}(r) = {1 \over r^\Delta} \left({r_+^2 \over r^2} - 1\right)^{-i \hw / 2}
{\Gamma(\Delta) \Gamma(i \hw) \over \Gamma((\Delta/2) + i \hw^+) \Gamma((\Delta/2) + i \hw^-)} \\
&&\qquad\qquad\qquad
F\left({\Delta \over 2} - i \hw^+, {\Delta \over 2} - i \hw^-, 1 - i \hw, 1 - {r_+^2 \over r^2}\right)~.
\eeas
Then a complete set of $\left\lbrace{{\rm positive} \atop {\rm negative}}\right\rbrace$ frequency
Kruskal modes is given by
\bea
\label{KruskalModes}
&& \phi_R^\pm(t,r,\phi) = e^{-i \omega t} e^{i k \phi} f_{\omega k}(r) \\
\nonumber
&& \phi_F^\pm(t,r,\phi) = e^{-i \omega t} e^{i k \phi} \left(
g_{\omega k}(r) + e^{\mp \pi \hw} g_{-\omega,k}(r)\right) \\
\nonumber
&& \phi_L^\pm(t,r,\phi) = e^{\mp \pi \hw} e^{-i \omega t} e^{i k \phi} f_{\omega k}(r)
\eea
in the (right, future, left) Rindler wedges.  This means we can express the value of the field in the
future wedge in terms of data on the right and left boundaries, via
\beas
\phi_F(t,r,\phi) & = & \int d\omega dk \, {1 \over 4 \pi^2} g_{\omega k}(r)
\Biggl[ \int dt' d\phi' \, \bigl(e^{-i \omega (t - t')} e^{i k (\phi - \phi')} \phinotrR{}(t',\phi') \\
&& \hspace{4.8cm}
+ e^{-i \omega(-t + t')} e^{i k (\phi - \phi')} \phinotrL{}(t',\phi')\bigr)\Biggr]\,.
\eeas
(Recall that time is oriented oppositely on the two boundaries, so for $t = 0$ this expression is in fact
symmetric between the right and left boundaries.)
Switching the order of integration and performing the $\omega$ and $k$ integrals first gives
a formal representation of the Rindler smearing function, essentially as the Fourier
transform of $g_{\omega k}$.  However it is easy to check that $g_{\omega k}$ grows exponentially as
$k \rightarrow \pm \infty$.  So we are not justified in switching the order of integration and the Fourier
transform does not exist.  One can presumably make sense of the Rindler smearing function in this approach
by deforming the contours of integration as in \cite{hkll2}.  For points inside the horizon this should reproduce
the result (\ref{InsideRindler}) we obtained from Poincar\'e coordinates.

\section{Non-integer $\Delta$}\label{NonIntegerDelta}

In this appendix we work out the generalization of the Rindler smearing function (\ref{InsideRindler})
appropriate for arbitrary conformal dimension.

We first need to discuss the generalization of the antipodal map.  This is easiest to understand in
global coordinates, where the embedding coordinates of section \ref{AdS3prelims} are given by
\beas
&& U = R \cos \tau / \cos \rho \\
&& V = R \sin \tau / \cos \rho \\
&& X = R \cos \theta \tan \rho \\
&& Y = R \sin \theta \tan \rho
\eeas
for $-\infty < \tau < \infty$, $0 \leq \rho < \pi/2$, $\theta \sim \theta + 2 \pi$.  The induced metric is
\[
ds^2 = {R^2 \over \cos^2 \rho} \left(-d\tau^2 + d\rho^2 + \sin^2 \rho d\theta^2\right)\,.
\]
The antipodal map acts by
\[
A \, : \, (\tau,\rho,\theta) \rightarrow (\tau - \pi,\, \rho, \theta + \pi)\,.
\]
The global mode expansion is
\[
\phi(\tau,\rho,\theta) = \sum_{n = 0}^\infty \sum_{l = -\infty}^\infty a_{nl} e^{-i \omega_{nl} \tau}
e^{i l \theta} \sin^{\vert l \vert} \rho \, \cos^\Delta \rho \, P_n^{(\vert l \vert,\Delta - 1)}(\cos 2 \rho) + c.c.
\]
where $\omega_{nl} = 2n + \vert l \vert + \Delta$ and $P_n$ is a Jacobi polynomial.  So
fields which are $\left\lbrace{{\rm positive} \atop {\rm negative}}\right\rbrace$ frequency with respect to global time
satisfy
\[
\phi^\pm(x) = e^{\mp i \pi \Delta} \phi^\pm(Ax)\,.
\]
This means the generalization of (\ref{InsideRindler}) to arbitrary conformal dimension is
\bea
& & \phi = {(\Delta - 1) 2^{\Delta - 2} \over \pi r_+^2} \biggl[\int_{\sigma > 0}
\hspace{-0.5cm} dx dy \,
\lim_{r' \rightarrow \infty} (\sigma / r')^{\Delta - 2} \phinotrR{}(\htt + x, \hp + i y) \\
\nonumber
& & + \int_{\sigma < 0} \hspace{-0.5cm} dx dy \,
\lim_{r' \rightarrow \infty} (-\sigma / r')^{\Delta - 2} \left(
e^{-i\pi\Delta}\phinotrL{+}(\htt + x, \hp + i y) + e^{i\pi\Delta}\phinotrL{-}(\htt + x, \hp + i y)\right)\biggr]
\eea
where we've decomposed the left boundary field into pieces $\phinotrL{\pm}$ that are $\left\lbrace{{\rm positive}
\atop {\rm negative}}\right\rbrace$ frequency with respect to global (equivalently, Kruskal) time. These may in turn be
expressed in terms of integrals involving $\phinotrL{}$ and $\phinotrR{}$ over all time, as in appendix B of \cite{hkll}.

\bibliographystyle{brownphys}
\bibliography{smear}

\end{document}